# Periodic Fluorescence Variations of CdSe Quantum Dots Coupled to Aryleneethynylenes with Aggregation Induced Emission

*Krishan Kumar[1], Jonas Hiller[1], Markus Bender[2], Saeed Nosrati[1], Quan Liu[1,3], Frank Wackenhut[1], Alfred J. Meixner[1,4], Kai Braun[1], Uwe H. F. Bunz[2,\*], Marcus Scheele[1,4,\*]*

[1] Institute for Physical and Theoretical Chemistry, University of Tübingen, Auf der Morgenstelle 18, 72076 Tübingen, Germany.

[2] Organisch-Chemisches Institut and Centre for Advanced Materials, Ruprecht-Karls-Universität Heidelberg, Im Neuenheimer Feld 270, 69120 Heidelberg, Germany

[3] Charles Delaunay Institute, CNRS Light, nanomaterials, nanotechnologies (L2n, former "LNIO") University of Technology of Troyes, 12 rue Marie Curie - CS 42060, 10004 Troyes Cedex, France

[4] Center for Light-Matter Interaction, Sensors & Analytics LISA+, University of Tübingen, Auf der Morgenstelle 15, 72076 Tübingen, Germany

**Abstract**

CdSe nanocrystals and aggregates of an aryleneethynylene derivative are assembled into a hybrid thin film with dual fluorescence from both fluorophores. Under continuous excitation, the nanocrystals and the molecules exhibit anti-correlated fluorescence intensity variations, which become periodic at low temperature. We attribute this to a structure-dependent aggregation induced emission of the aryleneethynylene derivative, which impacts the rate of excitation energy transfer between the molecules and nanocrystals. Energy transfer also affects the electric transport properties of the hybrid material under optical excitation. This work highlights that combining semiconductor nanocrystals with molecular aggregates, which exhibit aggregation induced emission, can result in unprecedented emerging optical properties.

**Introduction**

Aggregation induced emission (AIE) refers to enhanced and often redshifted fluorescence of luminophores upon formation of aggregates from solution.[1] The prototypical example are organic π-systems with a large degree of structural twisting in the ground state, e.g. tetraphenylethylenes[2] or aryleneethynylenes[3], which exhibit high torsion angles between the central C=C or C≡C bond and the sterically demanding phenyl-substituents. Photoexcitation weakens the double or triple bond and reduces the torsion angles with the phenyl rings.[4–6] Relaxation from this state occurs via three pathways: 1) radiative recombination, 2) intramolecular motion (mostly de-twisting of the C=C or C≡C bond) or 3) photochemical reaction to an intermediate. AIE occurs if pathways 2)+3) are significantly inhibited in the aggregates due to restriction of intramolecular motion.[7] The degree of this restriction depends on the solid-state structure and, thus, on temperature-induced structural transitions.[5,8,9] These transitions may be triggered photothermally via sufficiently strong photoexcitation and the concomitant rise in temperature.[10,11] The photoexcitation can occur either directly via resonant excitation of the aggregates or indirectly by excitation energy transfer (EET) via a sensitizer.[12,13] Inorganic semiconductor nanocrystals (NCs) are ideal as sensitizers as they exhibit much larger extinction coefficients than most organic dyes and strong absorption at above-band gap photon energies.[14] EET between NCs and organic π-systems has been extensively investigated in both: solution and solid state. However, to the best of our knowledge, no research has been committed specifically towards involving AIE molecules.[15–21]

Here, we combine aggregates of the aryleneethynylene derivative AE **1** (**Figure 1a**) with pronounced AIE together with CdSe NCs into a hybrid nanocomposite. We show that EET between the NCs and the molecular aggregates in combination with photothermally induced structural changes lead to temporal fluctuations of the NC fluorescence and the degree of AIE in the aggregates, which are anticorrelated to each other. At low temperature, these intensity fluctuations become periodic. EET plays a role in charge carrier transport during photocurrent experiments.

**Results**

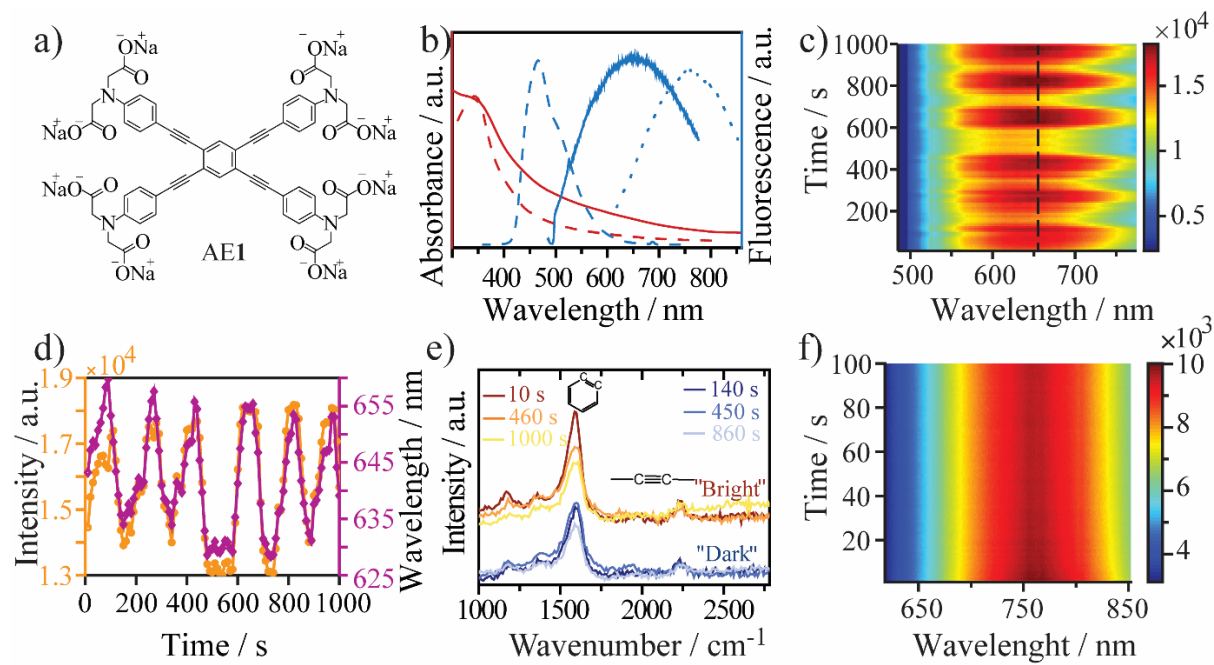

**Figure 1. a)** Structural formula of the arylenephenylene derivative AE **1**. **b)** Optical properties of AE **1**. Absorption in methanol (red dashed line), thin-film absorption (red solid line), fluorescence in methanol (blue dashed line, excitation at 350 nm), thin-film fluorescence (blue solid line, 488 nm excitation) and thin-film fluorescence at 160 K (blue dotted line, 488 nm excitation). **c)** Thin-film fluorescence during 1000 s of continuous excitation at 488 nm with a binning time of 1 s. **d)** Line profile (orange solid) and wavelength maxima position (purple) of c) extracted by fitting gaussian functions. **e)** Raman spectra taken during "bright" periods (at 10, 460 and 1000 s) and "dark" periods (at 140, 450 and 860 s). Bands attributed to the phenyl-C=C breathing and the C≡C stretching mode are indicated. **f)** Thin film fluorescence at 160 K during 100 s of continuous excitation at 488 nm.

In methanolic solution, AE **1** shows broad absorption with a maximum at 350 nm, including a weak low-energy tail, and narrow, dim emission at 480 nm (**Figure 1b, red and blue dashed lines, respectively**). In the solid state, aggregates of AE **1** exhibit pronounced low-energy tailing of the absorption as well as broad, enhanced emission with a maximum at 650 nm and a lifetime (τ) of 0.8 ns (**Figure 1b, red and blue solid lines**). At 160 K, the fluorescence maximum is further red-shifted to 750 nm (**Figure 1b, dotted line**). These optical properties are typical for molecules with AIE and are consistent with other, similar aryleneethynylenes.[6,8,22,23] Absorption and fluorescence of AE **1** are strongly dependent on the extent of conjugation of the π-electron cloud, for which the torsion angles of the five phenyl rings are an important measure.[24] For torsion angles close to 0 °, the molecule is fully planarized, the degree of π-conjugation is maximized and the energy of this rotamer is minimized.[25] In the ground state, the energy difference between rotamers of different torsion angles is low (< 4 KJ/mol) and a wide range of rotational states are populated.[26] This leads to the broad absorption feature in **Figure 1b**. In the excited state, the torsion-angle dependent energy profile of the rotamers becomes much steeper ($\Delta E \approx 30$ KJ/mol)[27] and, an unhindered rotation provided, radiative emission will occur from a narrow range of rotational states with relatively small torsion angles.[26] This narrows and red-shifts the fluorescence compared to the absorption. In the solid state, structural rigidity favors rotamers with small torsion angles and enforces a higher degree of planarization. This invokes the low-energy tailing of the absorption and the red-shift of the fluorescence in **Figure 1b**. Kinetic arrest of high energy rotamers due to the structural rigidity prevents the fluorescence line narrowing observed in solution and leads to a broad fluorescence feature. At low temperature, structural order and, thus, the degree of planarization, are enhanced further, which shifts the fluorescence to even lower energies.[28,29]

Continuous excitation at 488 nm under nitrogen atmosphere leads to substantial fluctuations (± 20 %) in the fluorescence intensity emitted by the AE **1** aggregates (**Figure 1c,d**). The timescale of seconds for the transition between an intensity maximum and minimum suggests that a macroscopic process, and not an electronic transition, is responsible for the fluctuations. Within the theory of AIE, this process needs to affect either the restriction of intramolecular motion or the rate of formation of a photochemical intermediate to result in fluorescence intensity fluctuations.[4] Photothermally induced local structural changes are a likely cause for less restricted intramolecular motion and faster non-radiative recombination. Indeed, we observe significant local changes in the optical scattering and luminescence images of AE **1** aggregates after laser excitation, indicating an altered morphology (see **Figure S1**). Furthermore, we find a strong correlation between the intensity and the peak wavelength of the fluorescence, in that a lower intensity coincides with a blue-shifted fluorescence peak (**Figure 1d**). This is consistent with a transformation from a more planarized structure with high fluorescence quantum yield to a less-planarized structure with fast non-radiative recombination. To test for the formation of a photochemical intermediate as an alternative cause for the intensity fluctuations, we compare the Raman bands of AE **1** appearing together with the fluorescence signals of different intensity (**Figure 1e**). We find a strong band at 1590 cm$^{-1}$ and a weak signal at 2230 cm$^{-1}$, which we assign to the phenyl-C=C breathing and the C≡C stretching mode in agreement with earlier reports.[29,30] There are no substantial differences between Raman spectra taken during periods of fluorescence intensity minima vs. maxima, suggesting that chemical transformations of AE **1** are not responsible for the fluorescence intensity fluctuations. Over the course of 1000 s of continuous excitation, the intensity of the phenyl-C=C breathing mode decreases in comparison with that of the C≡C stretching mode. However, this evolution progresses within periods of low and high fluorescence intensity alike.

We conclude that the fluorescence intensity fluctuations of AE **1** under continuous 488 nm laser excitation are probably caused by a photothermally induced order/disorder transition. This transition locally increases the non-radiative recombination rate due to bond rotation and decreases the fluorescence quantum yield. In the disordered, "solution-like" state, the absorption of AE **1** at 488 nm is weakened (**Figure 1b**), which now decreases the rate of photothermal heating and allows for a recovery of the initial optical properties of the AE **1** aggregates. At 160 K, the fluorescence fluctuations disappear (**Figure 1f**). We suspect that photothermal heating may not be sufficient at this temperature to invoke the same structural changes that lead to the random fluorescence fluctuations in **Figure 1c**.

In **Figure 2**, we analyze the solid-state fluorescence of AE **1** aggregates coupled to the surface of iodide-capped CdSe nanocrystals (CdSe/I$^-$ NCs) at room temperature. The NCs were chosen based on the large spectral overlap between their fluorescence (**Figure 2a, blue solid line**) with the absorption of AE **1** aggregates (**Figure 2a, red dashed line**) to enable efficient EET. The absorption spectrum of the hybrid material (CdSe/I$^-$/AE **1**) is dominated by the absorption of the CdSe NCs (**purple dashed line**), while the fluorescence spectrum bears two well-resolved, narrow bands at 636 nm and 734 nm (**purple solid line**). Wavelength-selective fluorescence lifetime measurements of these bands reveal τ = 0.8 ns at 734 nm and a longer lifetime at 636 nm with τ = 1.5 ns (**Figure 2b**). Based on this, we assign the band at 636 nm to the CdSe/I$^-$ NCs and the band at 734 nm to AE **1**.

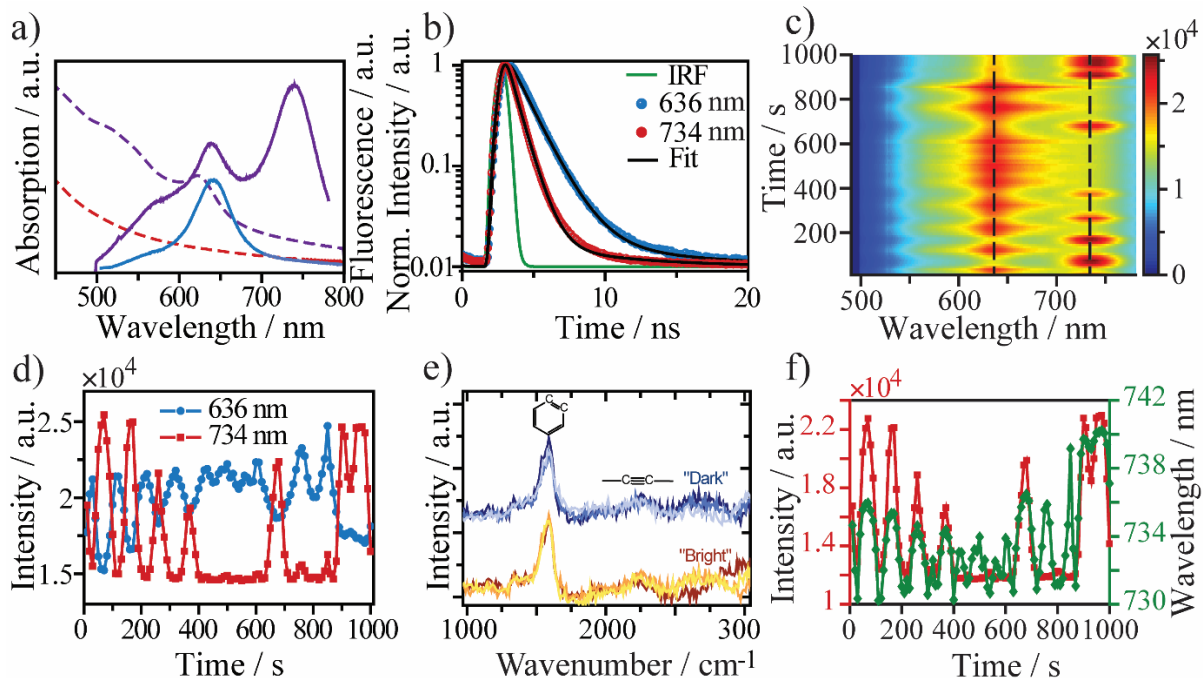

**Figure 2. a)** Thin-film absorption (purple dashed line) and thin-film fluorescence (purple solid line) of CdSe/I$^-$/AE **1**. Excitation at 488 nm. For comparison, the fluorescence of the CdSe/I$^-$ NCs (blue solid line) and the absorption of pure AE **1** (red dashed line) are also displayed. **b)** Fluorescence lifetime measurements of the CdSe/I$^-$/AE **1** bands centered around 636 nm (blue) and 734 nm (red) and their fit (black). The instrument response function (IRF) is displayed in green. **c)** Thin-film fluorescence of CdSe/I$^-$/AE **1** during 1000 s of continuous excitation at 488 nm with a binning time of 1 s. **d)** Line profile of c) cut at 636 nm and 734 nm. **e)** Raman spectra taken during "bright" and "dark" periods. **f)** Peak position of the low energy band in c) (green) compared with the intensity of the same band (red).

Continuous excitation at 488 nm with 1 to 10 MW/cm$^{-2}$ and nitrogen atmosphere of CdSe/I$^-$/AE **1** results in the temporal fluctuations in the fluorescence spectrum depicted in **Figure 2c**. The fluctuations occur gradually over a timescale of 10 – 30 s with ON/OFF-periods of similar duration. Most notably, the ON/OFF periods of the bands at 636 nm and 734 nm are mostly anti-correlated, that is, an increase of the NC fluorescence occurs together with a decrease of the AE **1** emission and vice versa (**Figure 2d**). We verify that CdSe/I$^-$ NCs without AE **1** do not exhibit similar fluorescence fluctuations (**Figure S2**). Similar to pure AE **1**, we find the same Raman signals at 1590 cm$^{-1}$ and 2230 cm$^{-1}$, which do not change significantly during relative ON or OFF periods (**Figure 2e**). Here however, the intensity of the phenyl-C=C breathing mode remains constant, indicating a higher photostability of the hybrid composite compared to pure AE **1**. Another similarity between CdSe/I$^-$/AE **1** and the pure molecule is the strong correlation of the intensity (red) and peak wavelength position (green) of the fluorescence of AE **1** (**Figure 2f**). In contrast, the peak wavelength position of the CdSe/I$^-$ fluorescence remains constant throughout the fluctuations.

At 160 K and helium atmosphere, the temporal fluorescence fluctuations of CdSe/I$^-$/AE **1** become periodic (**Figure 3**). At 45 kW/cm$^2$ excitation power density, the two fluorescence bands of the CdSe/I$^-$ NCs and the AE **1** aggregates are quasi-continuous over a timescale of 200 s (**Figure 3a**). Increasing the excitation power density to 615 kW/cm$^2$ induces periodic oscillations of the intensity of both fluorescence bands by ± 20 %, which remain exactly anti-correlated (**Figure 3b-e**). The peak wavelength position of AE **1** (**Figure 3d, purple**) follows the intensity fluctuations (**Figure 3d, blue**) almost perfectly, however, the variations are significantly smaller compared to room temperature (**Figure 2f**). The periodicity of the fluctuations depends on the excitation power density and varies with a period of 115 s at 80 kW/cm$^2$ to 26 s at 620 kW/cm$^2$ (**Figure 3f**) as determined from the sine fits shown in **Figure 3e**.

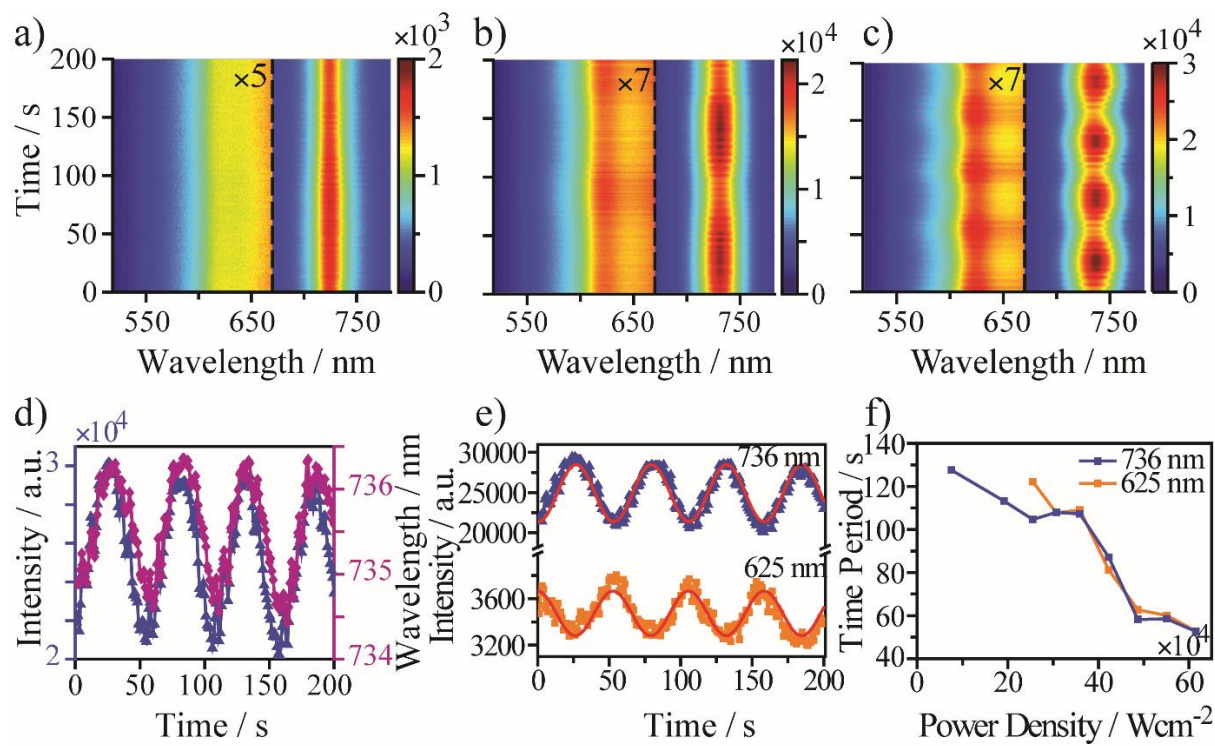

**Figure 3.** Thin-film fluorescence at 160 K of CdSe/I⁻/AE **1** during 200 s of continuous excitation at 488 nm with a binning time of 1 s and varying laser power. **a)** 45 kW/cm², **b)** 360 kW/cm² and **c)** 615 kW/cm². **d)** Peak position of the low energy band in c) (purple) compared with the intensity of the same band (blue). **e)** Line profile of c) cut at 625 nm and 736 nm. **f)** Correlation between the laser power and the period of one complete oscillation in e) for the fluorescence bands at 620 nm (blue) and 725 nm (orange).

Since thin films of CdSe/I⁻ NCs exhibit a strong field-effect and photocurrent response, we now explore the effect of including AE **1** into such devices.[31] **Figure 4a** compares the electric conductivity ($\sigma$) in the dark of CdSe/I⁻ NCs before (**blue**) and after (**orange**) functionalization with AE **1** between 8 – 300 K. The dark conductivities are mostly identical apart from a window of intermediate temperatures (100-200 K), where the presence of AE **1** invokes an increase in $\sigma$. Under optical excitation with $\lambda$ = 408 nm, the conductivities with AE **1** are much higher than for the bare CdSe/I⁻ NCs at all temperatures (**Figure 4b**). However, the corresponding field-effect mobilities under optical excitation (**Figure 4c & Figure S3**) are the same for both materials. Thus, the effect of AE **1** is mainly an increase of the free carrier concentration ($n$) in the NC ensemble under optical excitation, since $\sigma = \mu \cdot e \cdot n$, with the elemental charge $e$. Comparing the field-effect mobilities for CdSe/I⁻/AE **1** NCs in the dark (**Figure 4d, blue**) *vs.* the excited state (**orange**) reveals that transport under optical excitation becomes exceedingly more efficient than in the dark as the temperature decreases. This indicates the importance of the population of trap states: At low temperature, these states may only be populated through photoexcitation as thermalization is not sufficient anymore.

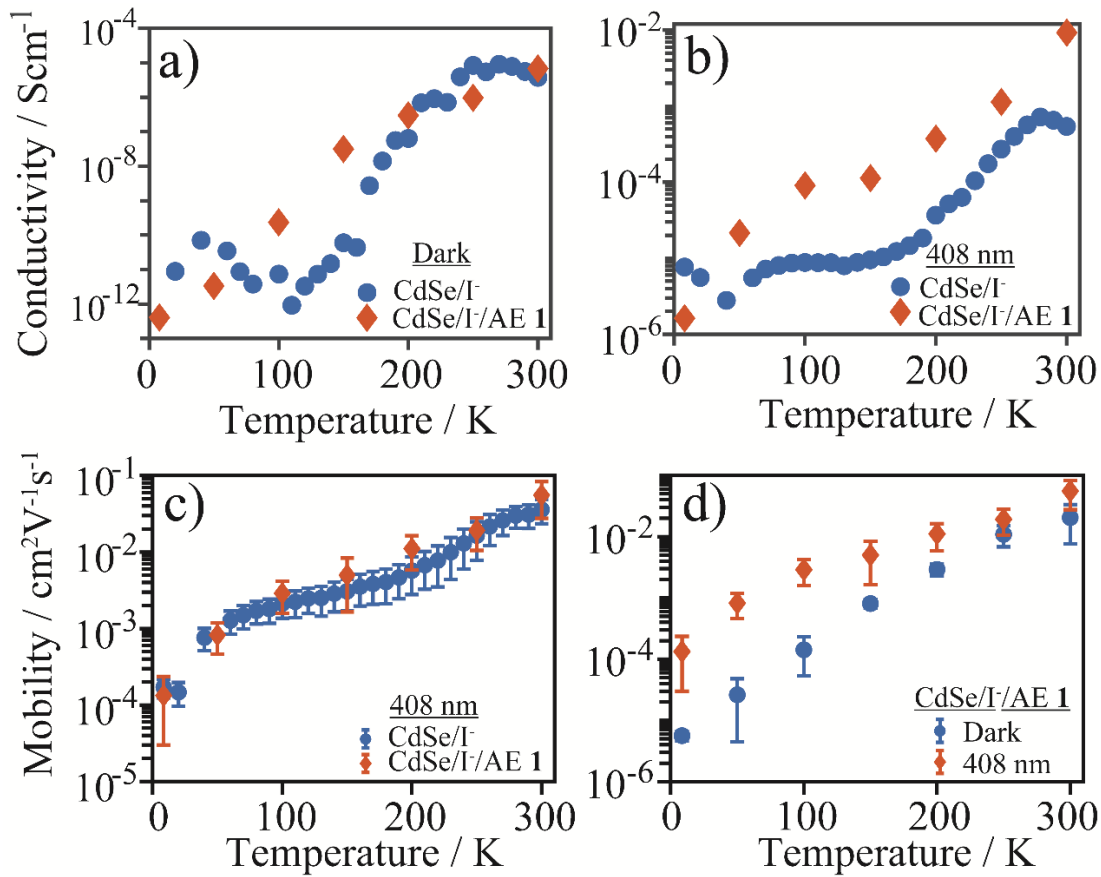

**Figure 4.** Temperature-dependent electric conductivity in the dark (**a**) and under 408 nm excitation (**b**) of thin films of CdSe/I⁻ NCs (blue) and CdSe/I-/AE **1** (orange). (**c**) Temperature dependence

under 408 nm excitation of the field-effect mobility of the same thin films. (**d**) Comparison of the dark (blue) and excited-state field-effect mobility (orange) of CdSe/I-/AE **1** thin films.

**Discussion**

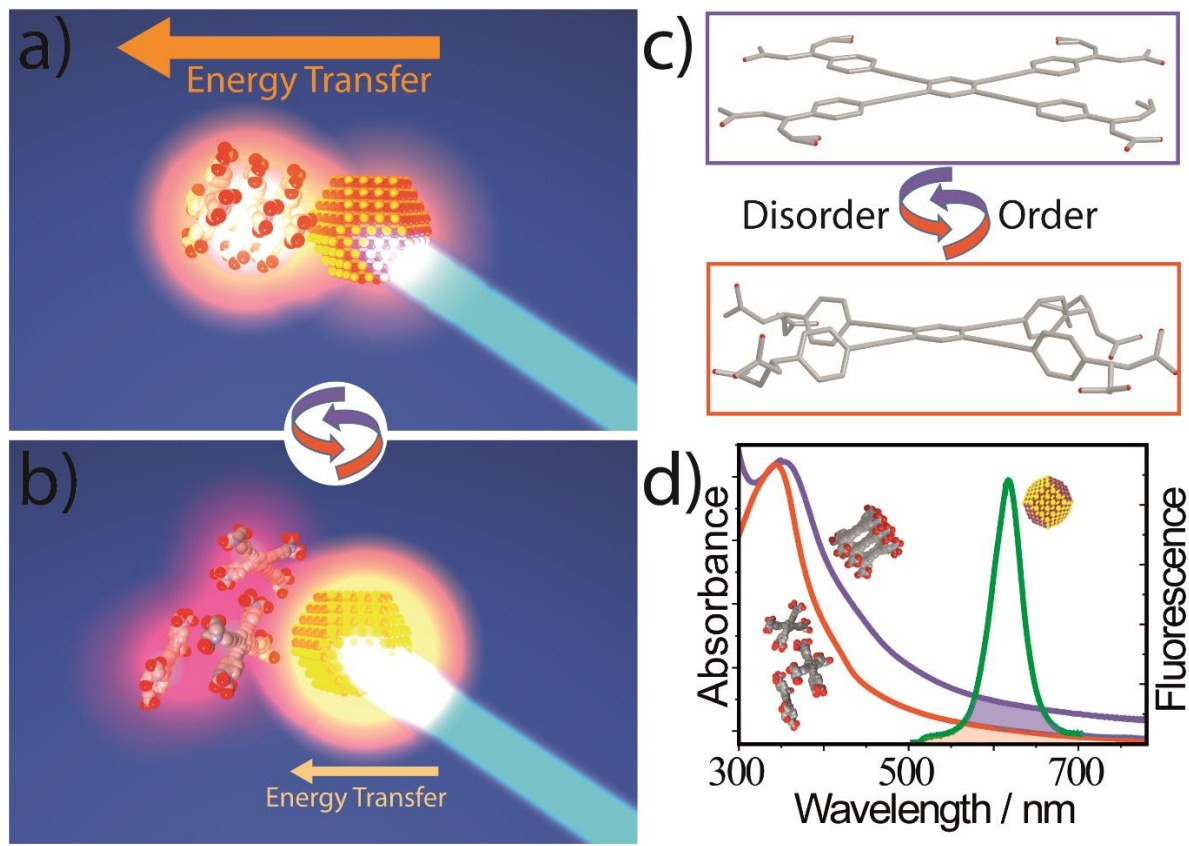

**Scheme 1**. **a-b).** Idealized schematic of the two structural scenarios proposed for CdSe/I$^-$/AE **1** composites with different energy transfer efficiency. **c)** Qualitative description of the proposed order/disorder transition. **d)** Spectral overlap between NC emission (green) and AE **1** absorption in the ordered (blue) and disordered (red) state.

We suggest that changes in the EET efficiency from excited CdSe/I$^-$ NCs to aggregates of AE **1** are responsible for the anticorrelated fluctuations in the CdSe/I$^-$/AE **1** nanocomposite (**Scheme 1a,b**). Under 488 nm excitation, most of the light is absorbed by the inorganic NCs as reflected by the absorption spectrum in **Figure 2a**, which mimics that of the pure NCs. The NC fluorescence overlaps with the absorption of AE **1**, resulting in efficient EET, which weakens the CdSe fluorescence signal and strengthens the molecular fluorescence. In **Figure 3e**, this scenario applies when the fluorescence band at 736 nm is at a maximum and the band at 625 nm is at a minimum, e.g. after 25 s. Simultaneously, this EET invokes photothermal heating of the AE **1** aggregates, eventually triggering an order/disorder transition to break up the aggregates. The result of this transition is a reduced degree of planarization (**Scheme 1c**), which blueshifts the absorption spectrum, decreases the overlap with the NC fluorescence (**Scheme 1d**), weakens the efficiency of EET, diminishes the molecular fluorescence and brightens the emission of the NCs. In **Figure 3e**, this scenario applies when the fluorescence band at 736 nm is at a minimum and the band at 625 nm is at a maximum, e.g. after 55 s. In the disordered state, EET is slow and photothermal heating is weak, such that new aggregates of AE **1** may form and the initial state with efficient EET from the NCs to the aggregates is restored. An illustration of the different absorption and emission pathways underlying this picture are provided in **Scheme S2** in the Supporting Information.

The overall interpretation in **Scheme 1** is consistent with the laser power-dependent period for one complete cycle (**Figure 3f**) as increasing the number of photons will increase the rate of heat transfer, which should shorten the cycle period. It is furthermore supported by the fluorescence fluctuations of pure AE **1** in **Figure 1**, which are likely due to a similar order/disorder transition of the aggregates. Relaxing the structural rigidity in molecular emitters with AIE increases the rate of non-radiative recombination and decreases the fluorescence intensity.[4] The higher stability of the Phenyl-C=C moiety in the nanocomposite (**Figure 2e**) compared to the pure molecule (**Figure 1e**) is consistent with mostly indirect excitation of AE **1** in the nanocomposite by EET *vs.* direct excitation of the pure molecule leading to photodegradation. EET as the main electronic interaction between the CdSe NCs and AE **1** is also consistent with the fluorescence lifetime data in **Figure 2b**. The essentially unchanged fluorescence lifetimes of the NCs and the molecular aggregates after combination into the hybrid material implies localized excitons, either on the NCs or the molecules. In particular, there are no indications for transfer of single charges and the formation of an interfacial exciton, for which one would expect much longer fluorescence lifetimes similar to those in type II core/shell NCs.[32]

The mobility data (**Figure 4c**) suggests that AE **1** does not enhance electronic coupling between the CdSe/I⁻ NCs. Transport occurs via hopping between adjacent NCs, irrespective of the presence of AE **1**. The increase in the carrier concentration with AE **1** under optical excitation (**Figure 4b**) implies that excitons dissociate into locally trapped charge carriers before they reach the electrodes, which mainly increases the free charge carrier density and not the field-effect mobility. A large trap state density is indicated by the differences in the dark vs. excited state mobilities at low temperature (**Figure 4d**). We suggest that reducing the density of trap states should increase the exciton diffusion lengths in CdSe/I⁻/AE **1** NC thin films with the potential to exploit EET as an additional means of charge carrier transport. This may enable similar periodic fluctuations in the electric transport properties as those demonstrated for the fluorescence.

**Conclusion**

Coupling organic π-systems that exhibit aggregation induced emission to a second fluorophore can result in emergent optical properties, such as dual fluorescence with oscillating intensity fluctuations. This is realized in a hybrid thin film consisting of CdSe nanocrystals coupled to aggregates of an aryleneethynylene derivative with fluorescence from both fluorophores, which establish a feedback loop: excitation energy transfer from the nanocrystals to the aggregates leads to photothermal heating and an order/disorder transition in the aggregates. This transition weakens the fluorescence intensity of the organic π-system and the rate of energy transfer, which strengthens the nanocrystal fluorescence. Simultaneously, the slower energy transfer rate reduces photothermal heating, which gradually restores the initial structure and optical properties of the molecular aggregates. This closes the feedback loop and initiates the next cycle. The resulting optical oscillations and their temperature dependence may be of interest for application in nanothermometry.

**Supporting Information**.

(**Figure S1**) Luminescence and scattering images of pure linker and CdSe/I-/AE 1, (**Figure S2**) Fluorescence of CdSe/I-, (**Figure S3**) Gate sweeps of CdSe/I- and CdSe/I-/AE 1. (**Scheme S1**) Schematic of time-resolved photoluminescence instrument. Experimental details and methods. (**Scheme S2**) Schematic of the absorption and emission pathways in CdSe/I-/AE 1 NC thin films in the ordered as well as the disordered state.

**Corresponding Author**

*Email: marcus.scheele@uni-tuebingen.de and uwe.bunz@oci.uni-heidelberg.de

**Author Contributions**

The manuscript was written through contributions of all authors. All authors have given approval to the final version of the manuscript.

**Acknowledgements**

Financial support of this work has been provided by the Emmy Noether program of the DFG under grant SCHE1905/3-1 and ME 1600/13-3.

# Supporting Information

# Periodic Fluorescence Variations of CdSe Quantum Dots Coupled to Aryleneethynylenes with Aggregation Induced Emission

*Krishan Kumar[1], Jonas Hiller[1], Markus Bender[2], Saeed Nosrati[1], Quan Liu[1,3], Frank Wackenhut[1], Alfred J. Meixner[1,4], Kai Braun[1], Uwe H. F. Bunz[2], Marcus Scheele[1,4]*

[1] Institute for Physical and Theoretical Chemistry, University of Tübingen, Auf der Morgenstelle 18, 72076 Tübingen, Germany.

[2] Organisch-Chemisches Institut and Centre for Advanced Materials, Ruprecht-Karls-Universität Heidelberg, Im Neuenheimer Feld 270, 69120 Heidelberg, Germany

[3] Charles Delaunay Institute, CNRS Light, nanomaterials, nanotechnologies (L2n, former "LNIO") University of Technology of Troyes, 12 rue Marie Curie - CS 42060, 10004 Troyes Cedex, France

[4] Center for Light-Matter Interaction, Sensors & Analytics LISA+, University of Tübingen, Auf der Morgenstelle 15, 72076 Tübingen, Germany

# AE 1 Synthesis

## General Remarks

All reactions requiring exclusion of oxygen and moisture were carried out in dry glassware under a dry and oxygen free nitrogen atmosphere. The addition of solvents or reagents was carried out using nitrogen flushed stainless steel cannulas and plastic syringes. For spectroscopic and analytic characterizations, the following devices were used:

**Analytical thin layer chromatography** (TLC) was performed on Macherey & Nagel Polygram® SIL G/UV254 precoated plastic sheets. Components were visualized by observation under UV light (254 nm or 365 nm) or in the case of UV-inactive substances by using the suitably coloring solutions. The following coloring solutions were used for the visualization of UV-inactive substances:

> $KMnO_4$ solution: 2.0 g $KMnO_4$, 10.0 g $K_2CO_3$, 0.3 g NaOH, 200 mL distilled water. Cer solution: 10.0 g $Ce(SO_4)_2$, 25 g phosphomolybdic acid hydrate, 1 L distilled water, 50 mL conc. $H_2SO_4$.

**Flash column chromatography** was carried out using silica gel S (32 μm-62 μm), purchased from Sigma Aldrich, according to G. Nill, unless otherwise stated.[1] As noted, Celite® 545, coarse, (Fluka) was used for filtration.

**Dialysis** was carried out using an appropriate length of the commercially available regenerated cellulose tubular membranes: Spectra/Por® Biotech Cellulose Ester (CE) Dialysis Membranes with the following specifications:

| | |
|---|---|
| Molecular weight cut off: | 500-1000 D |
| Flat width: | 31 mm |
| Vol/cm: | 3.1 mL/cm |
| Diameter in dry state: | 20 mm |

Unless stated otherwise the equipped tubular membranes were put into excess (~ 10 L) of deionized water and stirred for 5 d by changing the surrounding solvent once every day. The dialysed solution was freeze-dried afterwards.

**Melting points** (m.p.) were determined in open glass capillaries on a Melting Point Apparatus MEL-TEMP (Electrothermal, Rochford, UK) and are not corrected.

**¹H NMR** NMR spectra were recorded in CDCl$_3$ at room temperature on a Bruker DRX 300 (300 MHz), Bruker Avance III 300 (300 MHz), Bruker Avance III 400 (400 MHz), Bruker Avance III 500 (500 MHz) or Bruker Avance III 600 (600 MHz) spectrometer. The data were interpreted in first order spectra. All spectra were recorded in CDCl$_3$ or D$_2$O. Chemical shifts are reported in $\delta$ units relative to the solvent residual peak (CHCl$_3$ in CDCl$_3$ at $\delta_H$ = 7.27 ppm, HDO in D$_2$O at $\delta_H$ = 4.75 ppm or TMS ($\delta_H$ = 0.00 ppm).[2] The following abbreviations are used to indicate the signal multiplicity: s (singlet), d (doublet), t (triplet), q (quartet), quin (quintet), sext (sextet), dd (doublet of doublet), dt (doublet of triplet), ddd (doublet of doublet of doublet), etc., br. s (broad signal), m (multiplet), m$_c$ (centered multiplet). Coupling constants (*J*) are given in Hz and refer to H,H-couplings. All NMR spectra were integrated and processed using ACD/Spectrus Processor.

**¹³C NMR** spectra were recorded at room temperature on the following spectrometers: Bruker Avance III 400 (100 MHz) or Bruker Avance III 600 (150 MHz). The spectra were recorded in CDCl$_3$. Chemical shifts are reported in $\delta$ units relative to the solvent signal: CDCl$_3$ [$\delta_C$ = 77.00 ppm (central line of the triplet)] or TMS ($\delta_c$ = 0.00 ppm).

**High resolution mass spectra (HR-MS)** were either recorded on the JEOL JMS-700 (EI[+]), Bruker ApexQehybrid 9.4 T FT-ICR-MS (ESI[+]) or a Finnigan LCQ (ESI[+]) mass spectrometer at the Organisch-Chemisches Institut der Universität Heidelberg.

**Elemental Analyses** were carried out at the Organisch-Chemisches Institut der Universität Heidelberg.

**IR spectra** were recorded on a JASCO FT/IR-4100. Substances were applied as a film, solid or in solution. Processing of data was done using the software JASCO Spectra Manager™ II.

## Reagents and Solvents

Unless recorded otherwise, solvents were purchased from the chemical store of the Organisch-Chemisches Institut der Universität Heidelberg and distilled prior to use. Absolute solvents were prepared according to standard procedures and stored under argon over appropriately sized molecular sieve.[3] Absolute THF, Et$_2$O, toluene, CH$_2$Cl$_2$ and MeCN were purchased from Sigma-Aldrich and purified by a solvent purification system (MBraun, MB SPS-800, filter material: MB-KOL-A, MB-KOL-M; catalyst: MB-KOL-C). Chemicals other than solvents were either purchased from the chemical store of the Organisch-Chemisches Institut Universität Heidelberg or from commercial laboratory suppliers unless reported otherwise.

## Compounds

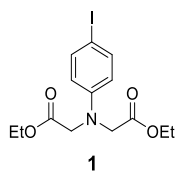

**Diethyl 2,2'-((4-iodophenyl)azanediyl)diacetate (1)**: Ethylbromoacetate (15.2 mL, 22.9 g, 137 mmol) was added to a solution of 4-Iodoanilline (10.0 g, 45.7 mmol) and *N*-ethyl-*N*-isopropylpropan-2-amine (40 mL) and was stirred at 120 °C for 4 h. The crude reaction mixture was quenched with $H_2O$, extracted with ethyl acetate, dried over $MgSO_4$ and evaporated *in vacuo*. The crude product was purified by flash chromatography on silica gel [petroleum ether/ethyl acetate (4/1)] to afford **1** (17.8 g, 99%) as an orange oil. $^1$H NMR (400 MHz, $CDCl_3$): δ = 6.48-6.38 (m, 4 H), 4.22 (q, *J* = 7.2 Hz, 4 H), 4.10 (s, 4H), 1.28 (t, *J* = 7.2 Hz, 6 H) ppm. $^{13}$C NMR (100 MHz, $CDCl_3$): δ = 170.4, 147.6, 137.8, 114.8, 79.8, 61.2, 53.5, 14.2 ppm ppm. IR ($cm^{-1}$): v 2979, 2935, 2905, 1730, 1588, 1561, 1495, 1445, 1430, 1370, 1343, 1316, 1294, 1255, 1175, 1113, 1095, 1076, 1053, 1022, 993, 968, 922, 855, 803, 727, 692, 631, 611, 550, 505, 441, 431. HR-MS (ESI$^+$): *m/z* calcd. for $C_{14}H_{19}INO_4^+$ 392.0353 [M+H]$^+$; found. 392.0355. $C_{14}H_{18}INO_4$ (391.03): calcd. C 42.98, H 4.64, N 3.58; found C 43.80, H 4.64, N 3.67.

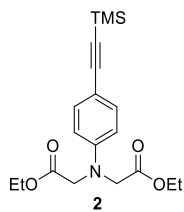

**Diethyl 2,2'-((4-((trimethylsilyl)ethynyl)phenyl)azanediyl)diacetate (2):** **1** (12.0 g, 30.7 mmol) was dissolved in THF/NEt$_3$ (2:1, 60 mL/30 mL) and degassed for 30 min with a stream of nitrogen. TMS-acetylene (5.68 mL, 3.92 g, 39.9 mmol), Pd(PPh$_3$)$_2$Cl$_2$ (393 mg, 613 µmol) and CuI (234 mg, 1.23 mmol) were added and the mixture was stirred under nitrogen at 50 °C for 48 h until TLC monitoring showed complete conversion. The reaction mixture was filtered over a pad of silica gel and evaporated *in vacuo*. The crude product was purified by flash chromatography on silica gel [petroleum ether/ethyl acetate (6/1)] to afford **2** (6.05 g, 68%) as a yellow solid (m. p. 53-55 °C). $^1$H NMR (300 MHz, $CDCl_3$): δ = 7.36-7.30 (m, 2 H), 6.55-6.50 (m, 2 H), 4.22 (q, *J* = 7.14 Hz, 4 H), 4.13 (s, 4 H), 1.28 (t, *J* = 7.14 Hz,

6 H), 0.23 (s, 9 H) ppm. $^{13}$C NMR (75 MHz, CDCl$_3$): δ = 170.4, 147.9, 133.3, 112.4, 112.0, 105.8, 91.9, 61.3, 53.4, 14.2, 0.1 ppm. IR (cm$^{-1}$): ν 2979, 2913, 2148, 1745, 1725, 1605, 1516, 1476, 1444, 1416, 1371, 1349, 1302, 1280, 1247, 1220, 1180, 1117, 1063, 1030, 968, 936, 858, 839, 820, 762, 737, 703, 650, 637, 599, 580, 534, 471, 449. HR-MS (ESI$^+$): *m/z* calcd. for C$_{19}$H$_{28}$NO$_4$Si$^+$ 362.1782 [M+H]$^+$; found 362.1784. C$_{19}$H$_{27}$NO$_4$Si (391.03): calcd. C 63.13, H 7.53, N 3.87; found C 63.24, H 7.62, N 3.76.

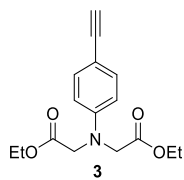

**Ethyl 2-(2,5-diethynyl-4-methoxyphenoxy)acetate (3)**: **2** (6.00 g, 16.6 mmol) was dissolved in THF (50 mL), placed in an ice bath and degassed for 15 min with a stream of nitrogen. TBAF (1M in THF, 18.4 mL) was added and the mixture was stirred for 10 min. The reaction mixture was quenched with SiO$_2$, filtered and evaporated *in vacuo*. The crude product was purified by flash chromatography on silica gel [(petroleum ether/ethyl acetate (6/1 → 1/1)] to afford **3** (2.54 g, 52%) as a light yellow/green oil. $^1$H NMR (300 MHz, CDCl$_3$): δ = 7.39-7.33 (m, 2 H), 6.58-6.51 (m, 2 H), 4.23 (q, *J* = 7.14 Hz, 4 H), 4.14 (s, 4 H), 2.97 (s, 1 H), 1.28 (t, *J* = 7.14 Hz, 6 H) ppm. $^{13}$C NMR (75 MHz, CDCl$_3$): δ = 170.4, 148.0, 133.4, 112.1, 111.3, 84.2, 75.3, 61.3, 53.4, 14.2 ppm. IR (cm$^{-1}$): ν 3280, 2981, 2937, 2101, 1731, 1607, 1557, 1515, 1446, 1388, 1371, 1343, 1324, 1298, 1257, 1175, 1114, 1095, 1056, 1022, 968, 932, 856, 816, 715, 646, 591, 533, 454, 438, 405. HR-MS (ESI$^+$): *m/z* calcd. For C$_{16}$H$_{20}$NO$_4$$^+$ 290.1387 [M+H]$^+$; found 290.1388. C$_{16}$H$_{19}$NO$_4$ (289.13): calcd. C 66.42, H 6.62, N 4.84; found C 66.23, H 6.78, N 4.68.

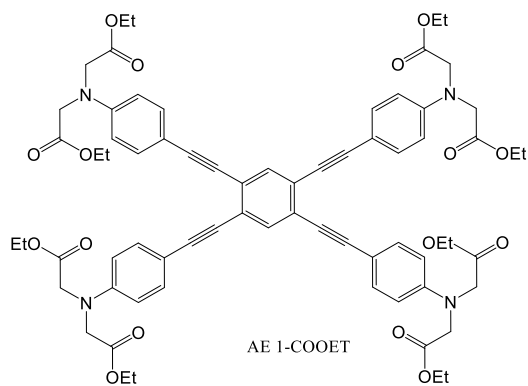

**Octaethyl 2,2',2'',2''',2'''',2''''',2'''''',2'''''''-(((benzene-1,2,4,5-tetrayltetrakis(ethyne-2,1-diyl))tetrakis(benzene-4,1-diyl))tetrakis(azanetriyl))octaacetate (AE 1-COOET)**: 1,2,4,5-Tetraiodobenzene[4] (320 mg, 550 µmol) and **3** (686 mg, 2.31 mmol) were dissolved in degassed THF/NEt$_3$ (1:1, 1.5 mL/1.5 mL). Pd(PPh$_3$)$_2$Cl$_2$ (39 mg, 55 µmol) and CuI (11 mg, 55 µmol) was added and the mixture was stirred under nitrogen at 60 °C for 24 h. The reaction mixture was filtered over a pad



of silica gel and evaporated *in vacuo*. The crude product was purified by flash chromatography on silica gel [petroleum ether/ethyl acetate (1/1)] to afford **AE 1-COOET** (398 mg, 59%) as a red film. $^1$H NMR (600 MHz, CDCl$_3$): δ = 7.64 (s, 2 H), 7.44-7.41 (m, 8 H), 6.61-6.58 (m, 8 H), 4.22 (q, *J* = 7.15 Hz, 18 H), 4.17 (s, 16 H), 1.30 (t, *J* = 7.15 Hz, 24 H) ppm. $^{13}$C NMR (150 MHz, CDCl$_3$): δ = 170.5, 147.9, 134.4, 133.0, 124.8, 112.5, 112.3, 95.6, 86.6, 61.3, 53.4, 14.2 ppm. IR (cm$^{-1}$): ν 2979, 2935, 2197, 1730, 1604, 1556, 1520, 1444, 1386, 1370, 1343, 1296, 1257, 1174, 1135, 1094, 1057, 1020, 967, 855, 813, 727, 647, 531, 467, 441. HR-MS (ESI$^+$): *m/z* calcd. for C$_{70}$H$_{74}$N$_4$NaO$_{16}$$^+$ 1249.4992 [M+Na]$^+$; found 1249.5004. C$_{70}$H$_{74}$N$_4$O$_{16}$ (1227.37): calcd. C, 68.50; H, 6.08; N, 4.56; found C 67.85, H 6.51, N 4.11.

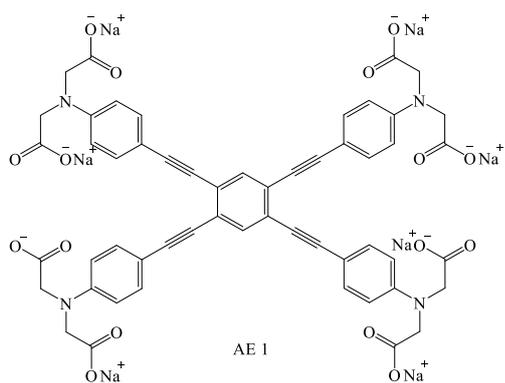

**Sodium 2,2',2'',2''',2'''',2''''',2'''''',2'''''''-(((benzene-1,2,4,5-tetrayltetrakis(ethyne-2,1-diyl))-tetrakis(benzene-4,1-diyl))tetrakis(azanetriyl))octaacetate (AE 1)**: **AE 1-COOET** (316 mg, 257 mmol) was dissolved in MeOH/H$_2$O (1:1, 10 mL/10 mL) and NaOH (515 mg, 12.9 mmol) was added. The resulting mixture was stirred at 70°C for 2 d. The solvent was evaporated *in vacuo.* The residue was dissolved in H$_2$O and filtrated. The resulting solution was adjusted to pH 7 and dialyzed against DI H$_2$O for 5 d. After filtration and freeze-drying the title compound **AE 1** was afforded as brown fluffy solid (250 mg, 97%). $^1$H NMR (600 MHz, D$_2$O): δ = 7.71-7.69 (m, 2 H), 7.48-7.44 (m, 8 H), 6.58-6.54 (m, 8 H), 3.93 (m, 16H) ppm. Due to low solubility, $^{13}$C NMR spectrum could not be obtained. IR (cm$^{-1}$): ν 3589, 3327, 3171, 3037, 2928, 2650, 2193, 1577, 1516, 1457, 1381, 1298, 1232, 1176, 1134, 972, 906, 817, 695, 610, 515, 456, 435.

# Selected $^1$H-NMR Spectra

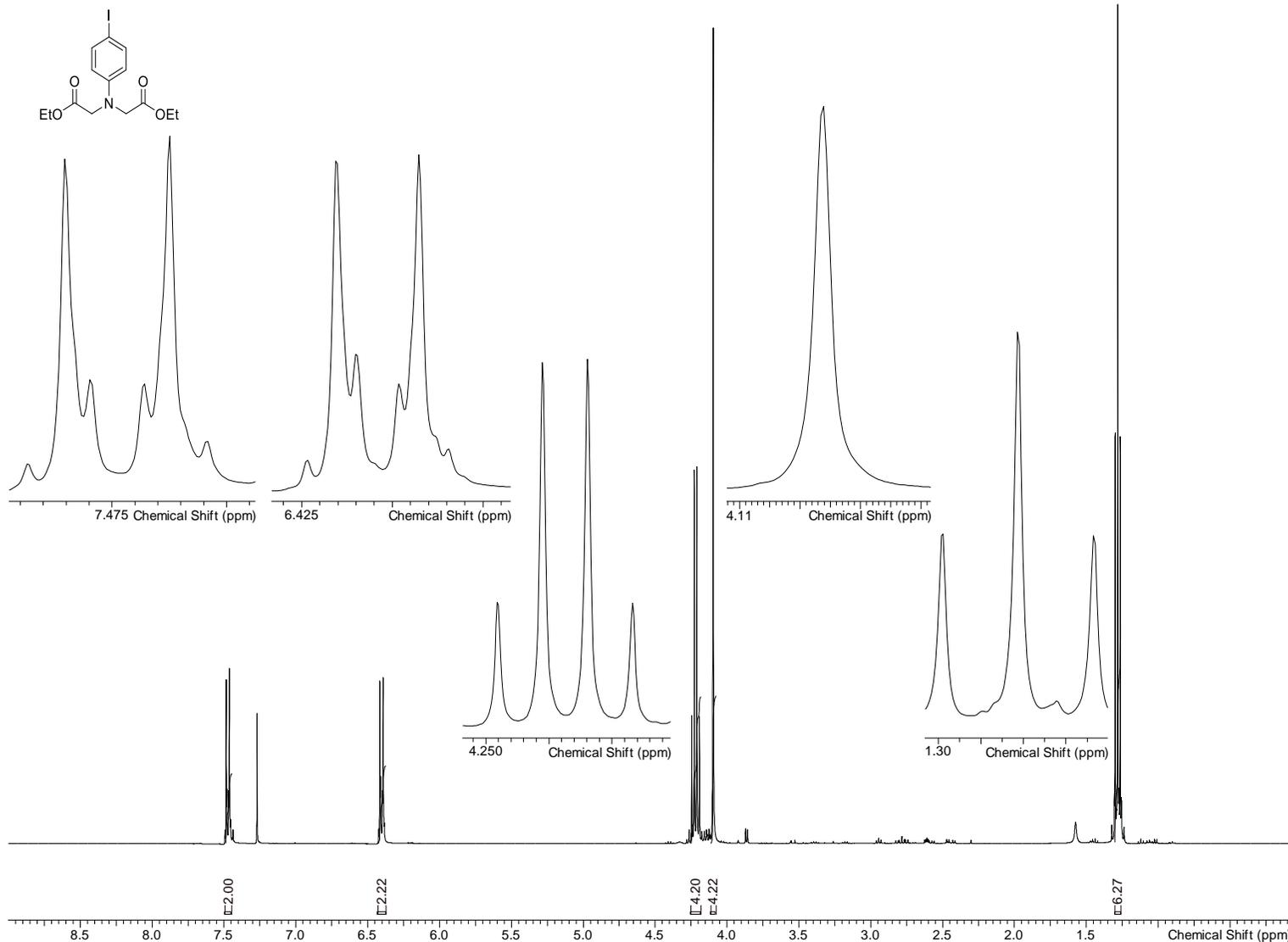

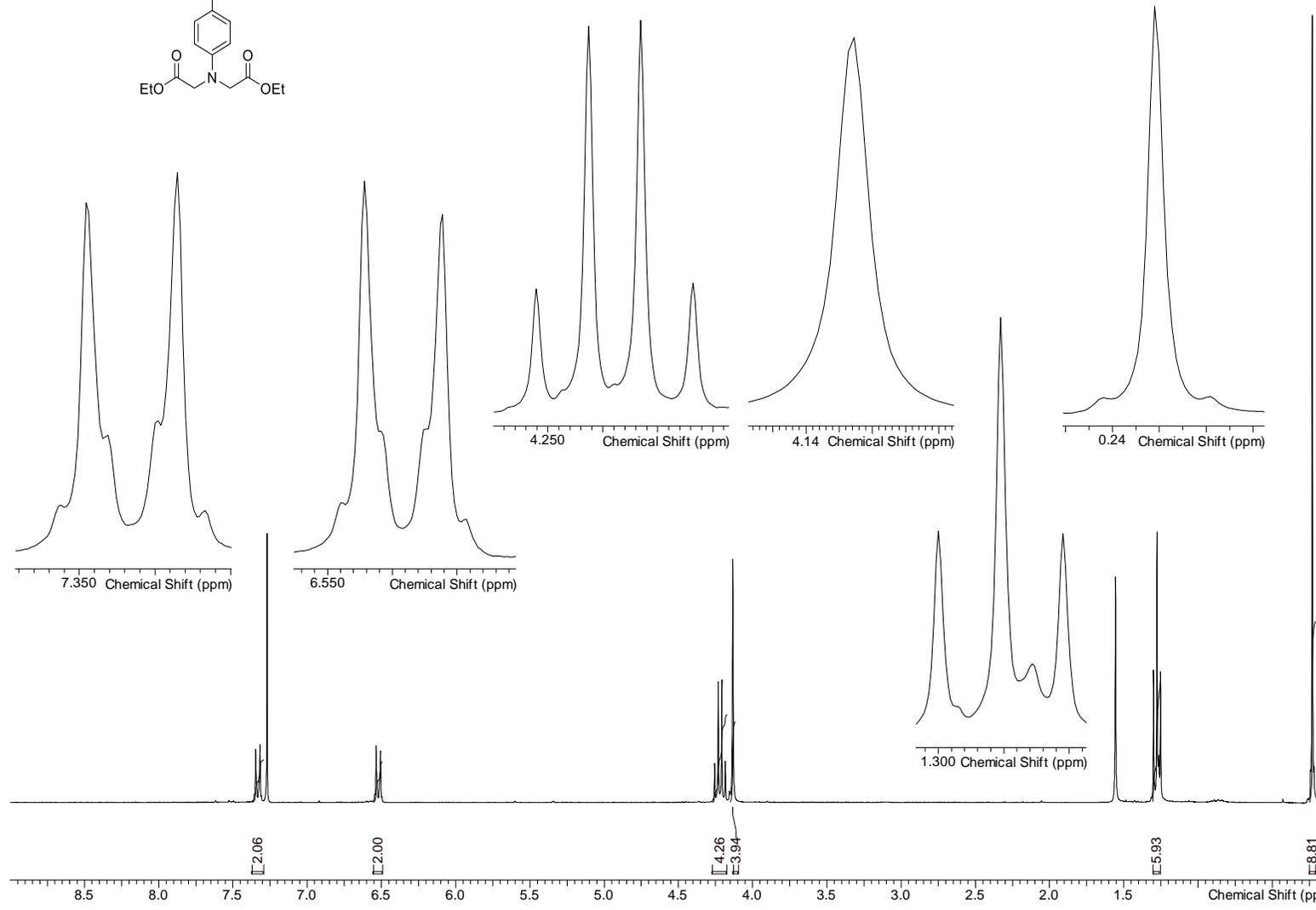

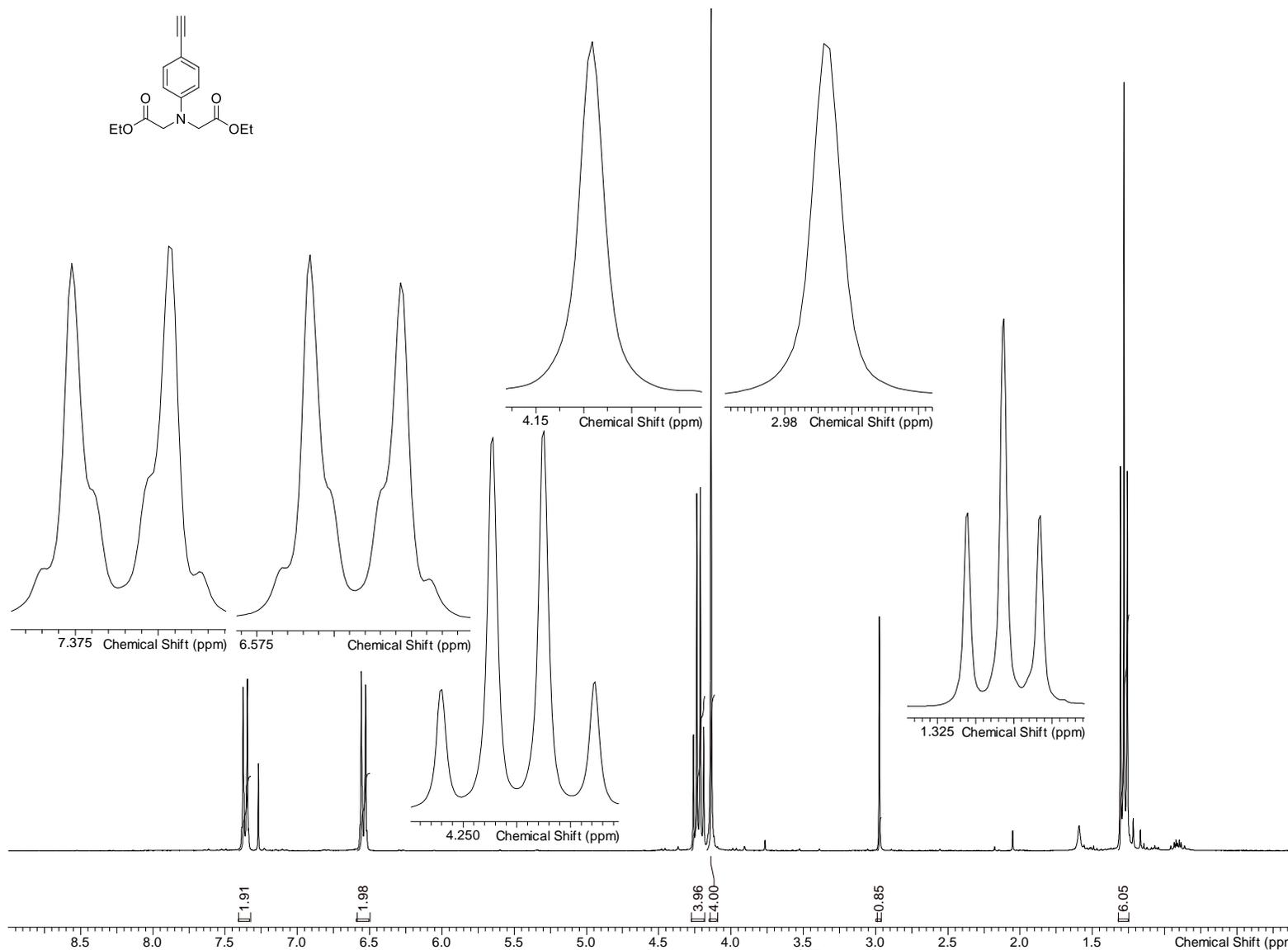

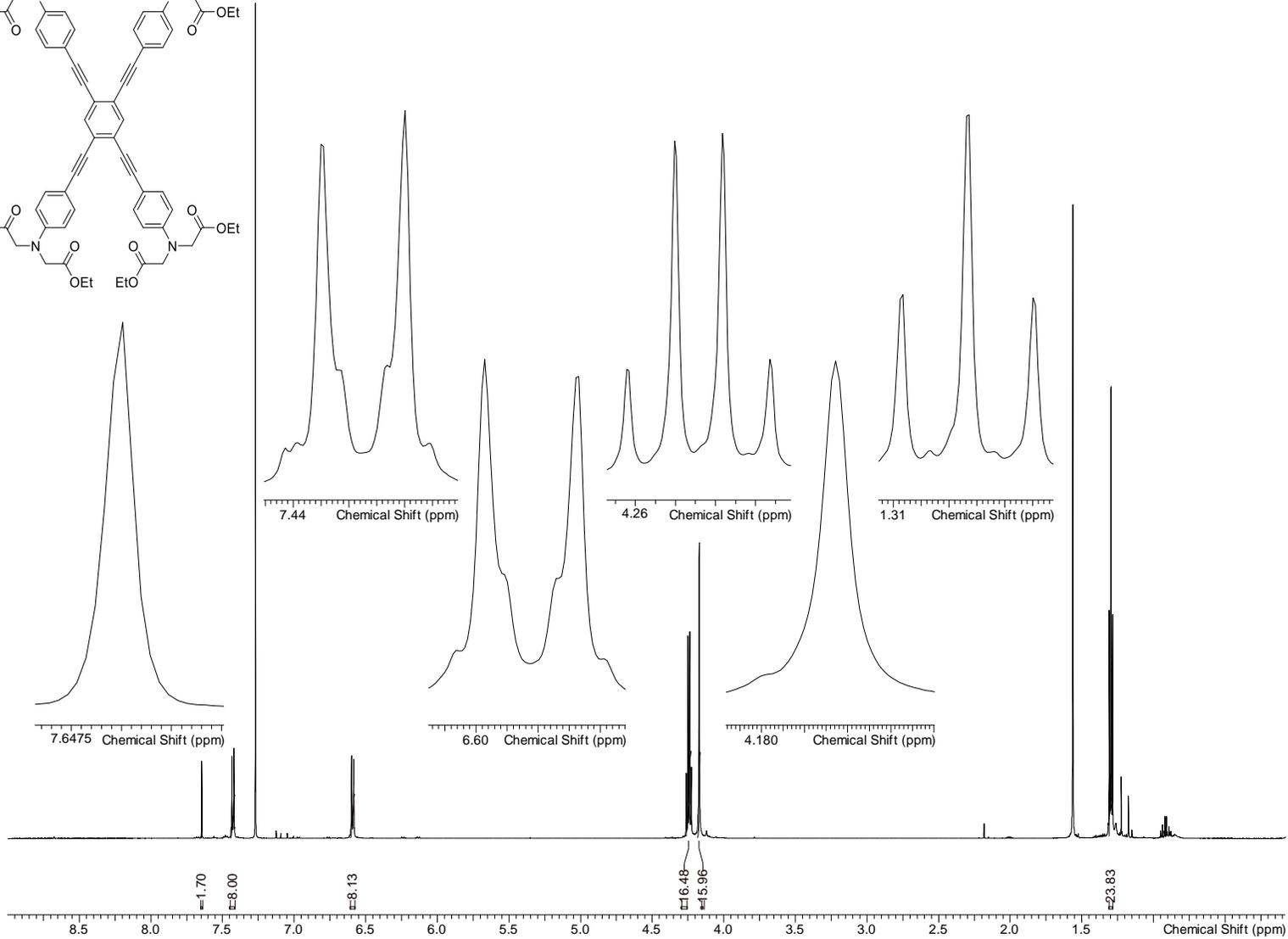

# CdSe NCs synthesis

## Chemicals Used

Chemicals used were cadmium oxide (CdO, 99.99%, Aldrich), oleicacid (OA, 90%, Aldrich), trioctylphosphine (TOP, 97%, Abcr), trioctylphosphine oxide (TOPO, 99%, Aldrich), hexadecylamine(HDA, 90%, Aldrich), 1-octadecene (ODE, 90%, Acros Organics), selenium pellet (Se, 99.999%, Aldrich), ammonium iodide (99.999%,Aldrich), N-methylformamide (NMF, 99%, Aldrich), hexane (ExtraDry, 96%, Acros Organics), ethanol (Extra Dry, 99.5%, AcrosOrganics), acetone (Extra Dry, 99.8%, Acros Organics), dimethylsulfoxide (DMSO, 99.7%, Acros Organics), and acetonitrile (ExtraDry, 99.9%, Acros Organics).

All chemicals, except those used in CdSe NC synthesis, were stored and used inside a nitrogen-filled glovebox. All sample preparations for electrical or fluorescence measurements were carried out in a nitrogen-filled glovebox. The samples were inserted into probe station for low temperature photocurrent measurements using an air tight arm sealed inside the glove box. Samples were kept under low pressure for at least 2 h before starting any measurements.

## CdSe NCs and Device Preparation

Wurtzite CdSe NCs having ~5 nm size were synthesized using a literature reported synthesis.[5,6] As synthesized NCs were dispersed in hexane. CdSe NCs 5 mL, ~10 mg/mL were taken for ligand exchange with 300 µL of a 1 M solution of $NH_4I$ in NMF further diluted using 2.7 mL acetone. This biphasic mixture was stirred until the NCs change their phase, then centrifuged and washed using excess of acetone. CdSe/I$^-$ thus obtained were dispersed in NMF having 60–100 mg/mL concentration. For device preparation we used commercially available gold patterned Si/SiO$_2$ with 230 nm thick dielectric layer and 2.5 $\mu m \times 1\ cm$ channel provided by the Fraunhofer Institute for Photonic Microsystems, Dresden, Germany. In a typical device preparation, 70 µL of CdSe/I$^-$ NCs were dropped on an FET substrate, mixed with 30 µL of AE **1** solution in NMF with a concentration of roughly 1 mg/mL and kept undisturbed for 6 h. The still wet mixture of CdSe/I$^-$/AE **1** was then spun off at 30 rps for 1 min. Then the substrate was washed with acetonitrile multiple times and annealed at 190° C for 35 min. A similar procedure was followed for preparing samples on coverslips for fluorescence measurements.

## Electrical and Optical Measurements:

Electrical measurements were carried out under nitrogen by using a *Keithley 2634B source meter*. The charge-carrier mobility was extracted using the gradual channel approximation in the linear regime. The error in the mobility (S) was calculated using the standard deviation error in the slope of $I_d$ vs $V_g$ curve at 5 V source-drain voltage:

$$S = \sqrt{\frac{1}{N-1}\sum_{i=1}^{N}|A_i - \overline{A}|}; \qquad \overline{A} = \frac{1}{N}\sum_{i=1}^{N} A_i$$

Here, A is the slope from N measurements with the mean value $\overline{A}$.

Absorption measurements were acquired using a *Carry 5000 UV-Vis-NIR* spectrophotometer on a thin glass coverslip or in methanol as stated in the text. Photocurrent low temperature measurements were recorded using a *CRX-6.5K (Lake shore Desert)* probe station operated under low pressure $5 \times 10^6\ mbar$ and a *Keithley 2634B source meter*. Samples were illuminated using a 408 nm *LP405-SF10* laser diode manufactured by *Thorlabs* having theoretical maximum output power 11.5 – 30 mW. This output power decreases orders of magnitude due to beam decollimation, scattering and inefficient coupling of optical fiber when calibrated using a test sample comes out to be 10 – 18 µW.

## Room Temperature Fluorescence Measurements

$12 \times 12\ mm$ glass coverslips were cleaned by submerging in a chromic acid cleaning solution for several hours followed by three subsequent washing steps with triple distilled water and spectroscopy grade methanol. Fluorescence samples were then prepared on these coverslips using the drop casting method described above. The film was washed with acetonitrile and annealed at 190 °C before taking any measurements.

The room temperature steady-state photoluminescence spectral measurements as well as photoluminescence and scattering image acquisition were carried out using a homebuilt inverted confocal laser scanning microscope.[7] A 488nm *TOPTICA Photonics* iBeam smart diode laser with a gaussian intensity profile, operated in continuous wave mode was employed as the light source. The laser intensity in the diffraction limited focus at maximum laser power is estimated to be $10^7\ W/cm^2$ operated at roughly 30-60% of maximum power. Focusing of the laser on the sample and the subsequent collection of reflected, as well as scattered and emitted light was achieved through an oil immersion objective (NA = 1.25). The spectral data was recorded using an *Acton SpectraPro 2300i* spectrometer with a grating of 300/mm and a detector temperature of -45 °C. Photoluminescence and scattering images were acquired by scanning the area of interest while utilizing two separate avalanche photodiodes (APDs) as detectors. The exclusion of atmospheric oxygen was achieved by nearly completely enclosing the upper part of the sample holder and passing a constant flow of nitrogen through this apparatus.

## Low Temperature Fluorescence Measurements

Low temperature fluorescence was measured using a home-built confocal microscope mounted on a damped optical table and a standard microscope objective (60X DIN Achromatic objective, NA = 0.85, Edmund Optics) located inside a cryostat (SVT-200, Janis). A Cernox temperature sensor (CX-1030-SD-HT 0.3L) was positioned close to the sample to measure the temperature by a LakeShore Model 336 temperature controller. The sample holder was mounted on the scan stage. Attocube systems linear stages (ANPx320 and ANPz101eXT) and scanners (ANSxy100lr and ANSz100lr) were used to scan and position the sample.

A continuous wave 488 nm laser diode (OBIS LS 20 mW, Coherent Inc) was used to excite the sample. The excitation intensity of the laser was measured between 0.35 mW and 4.80 mW before entering into the cryostat. The excitation light was then aligned into the objective (60X DIN Achromatic air objective, Edmund Optics) to get an optimal focus. The excitation intensity in the focus was between $5.5 \times 10^4$ and $7.52 \times 10^5 \ W/c m^2$.[8]

The collected fluorescence signal was passed through the dichroic mirror and a longpass filter (488 LP Edge Basic, AHF Analysentechnik). It was detected by a single-photon counting avalanche photodiode (APD, COUNT-100C, Laser Components). Fluorescence spectra were also acquired with integration times of one second by a Shamrock 500 spectrograph in combination with an Andor Newton back illuminated deep depleted CCD camera (DU920PPR-DD). Further details for low temperature confocal imaging and spectroscopy setup can be found elsewhere.[9]

## Time-resolved photoluminescence decay measurements

Time-resolved photoluminescence spectra were measured with a home built scanning confocal microscope.[10] The sample was fixed on a piezo stage (Physik Instrumente) via magnets to avoid movement. A constant nitrogen flow was applied to maintain an inert atmosphere and avoid oxidation of the sample. To avoid bleaching, the lifetime was always measured before a fluorescence measurement.

A linearly polarized continuous wave laser (488 nm, 0.33 mW measured before objective lens) was focused on the sample by a high numerical aperture (NA=1.46) oil objective, the fluorescence was collected by the same objective and sent to a spectrometer (Acton SP-2500i, Princeton Instruments). For lifetime measurement, the laser was operated in the pulsed mode ($5.3 \times 10^3\ W/cm^2, 20\ MHz$). The signal was sent to a single photon avalanche photodiode (APD), connected to a time-correlated Single Photon Counting module (TCSP, HydraHarp 400). Decay curves were fitted and analysed by SymPhoTime 64.

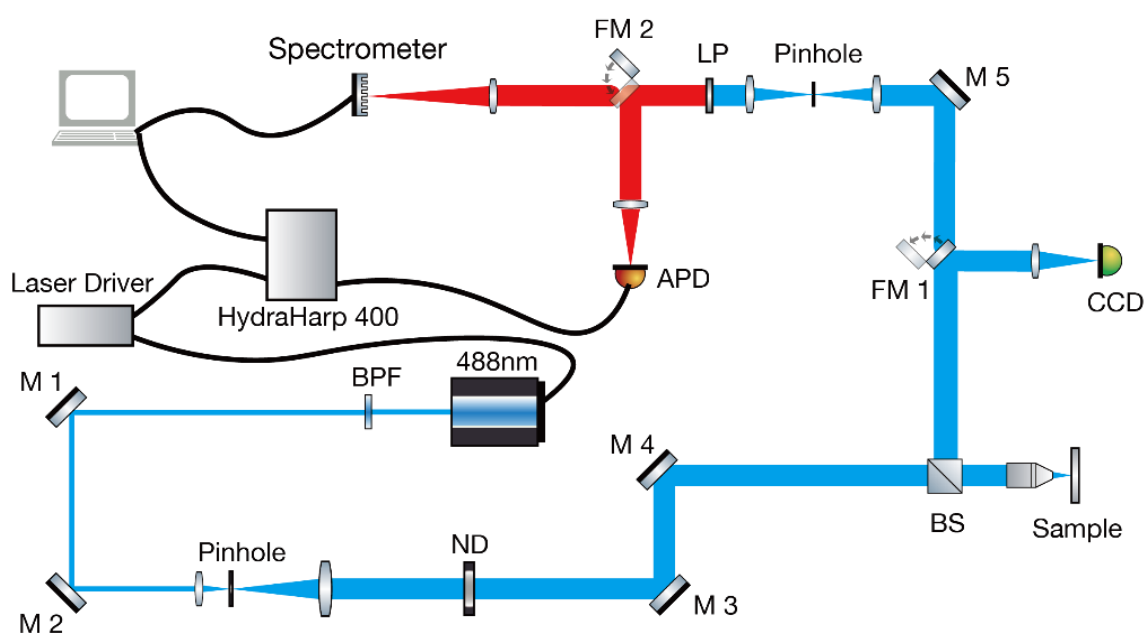

**Scheme S1.** Sketch of optical setup for fluorescence lifetime measurement. Abbreviations FM: Flip mirror, LP: Long-pass filter, M: Mirror, BPE: Band pass filter, ND: neutral-density filter, BS: Beam split.

# Supplementary Figures

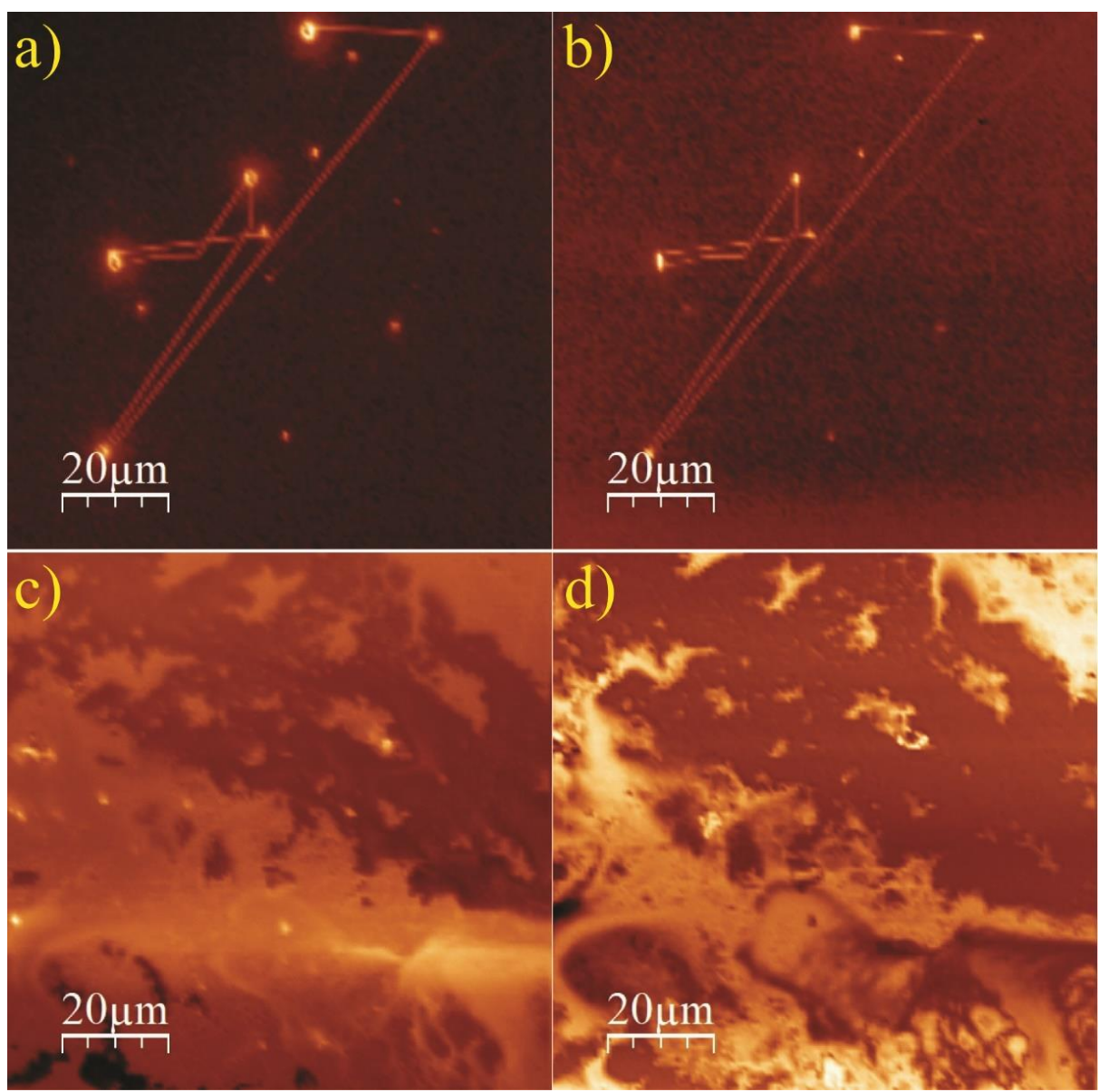

**Figure S1 a&c)** Luminescence and **b&d)** scattering images of pure linker (upper panel) & CdSe/I-/AE **1.** Pure linker films show laser tracing after continuous exposure to 488 nm laser indicating some changes in the film.

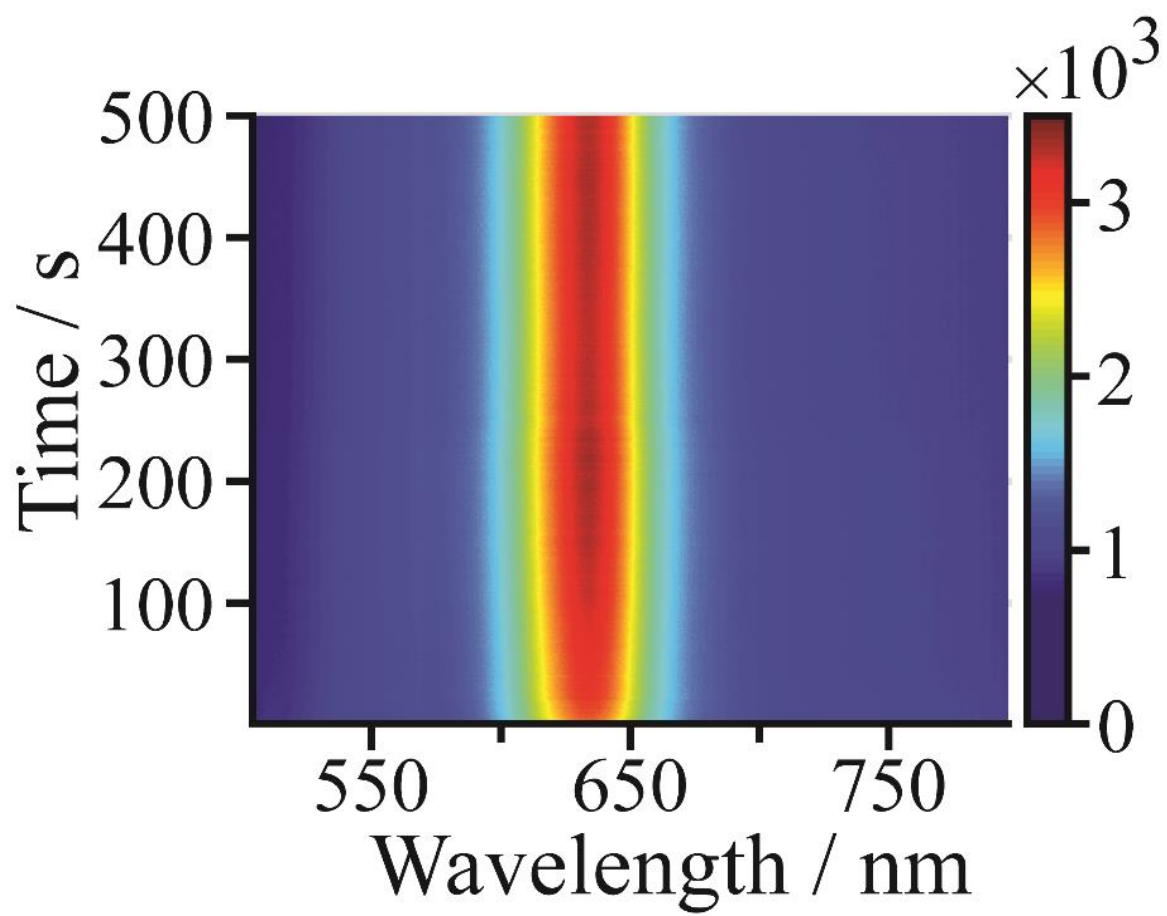

**Figure S2**. Thin film fluorescence of CdSe/I- NCs during 500 s of continuous excitation at 488 nm.

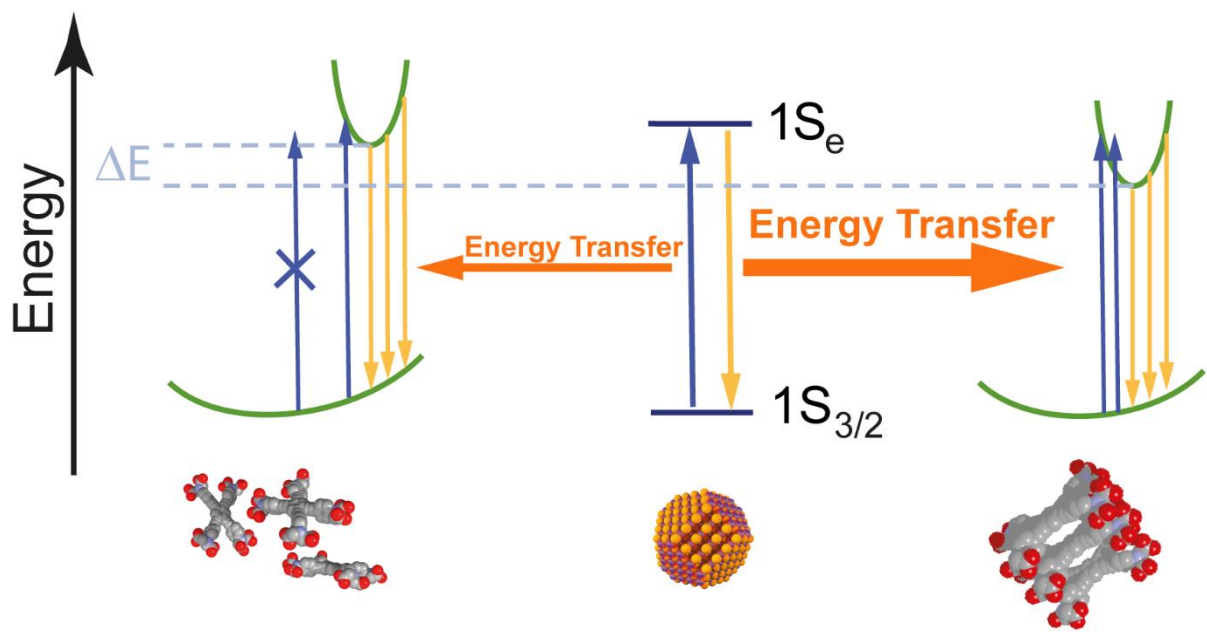

**Scheme S2.** Simplified energy level diagram of CdSe/I⁻/AE **1** NC thin films in the ordered state (**right**) as well as in the disordered state (**left**). **Blue arrows** indicate absorption, **yellow arrows** indicate emission pathways. Upon resonant excitation of the $1S_{3/2}$ - $1S_e$ transition of the NCs, the electron in the NC can either relax to the ground-state by radiative recombination or by transfer of its energy to AE **1** (**orange arrows**). For energy transfer, the emission energy in the NCs needs to match the absorption energy in AE **1**. This absorption energy exhibits a broad distribution due to the large number of rotamers with different energies (**green parabolas**), especially in the excited state, which are locked-in in the solid state. For ordered stacks of AE **1** molecules (**right**), there are several transitions between rotamers, which can be excited via energy transfer from the NCs. Thus, energy transfer is relatively efficient, resulting in bright fluorescence of AE **1**. In the disordered state (**left**), the average transition energy between rotamers is larger by ΔE compared to the ordered state. This is a consequence of the smaller degree of planarization of the molecules. In effect, the transferred energy from the NCs is not sufficient to excite most of the transitions between different rotamers in the disordered state, which leads to poor energy transfer and weak emission.

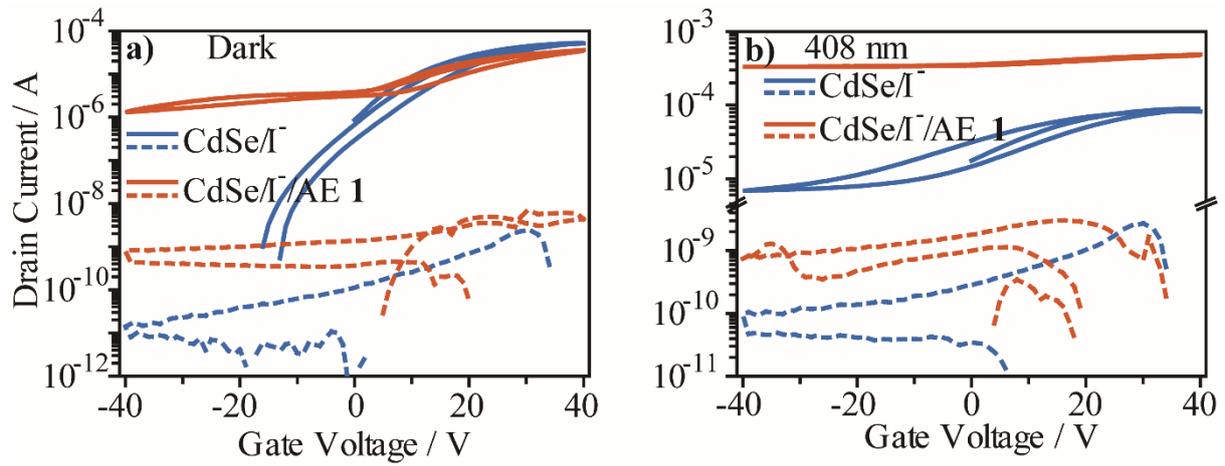

**Figure S3** Typical gate-sweep curve at 5 V source-drain voltage measured at 200 K with channel dimension 2.5 $\mu m \times 1\ cm$ **a)** under no illumination **b)** under 408 nm laser illumination. Solid curve represents the drain current and dotted curve represent the respective leakage current; negative data have been ignored for the logarithmic plot.